# TOWARDS A ROBUST QUALITY ASSURANCE FRAMEWORK FOR CLOUD COMPUTING ENVIRONMENTS

Mohammed Ahmad Alharbi   and Rizwan Qureshi

Department of Information Technology, Faculty of Computing and Information Technology, King Abdul-Aziz University, Jeddah 80213, Saudi Arabia

*ABSTRACT*

*Trends such as cloud computing raise issues regarding stable and uniform quality assurance (QA) and validation of software requirements. Current QA frameworks are poorly defined, often not automated, and lack the flexibility needed for on-demand, cloud-based environments. These gaps lead to inconsistencies in service delivery, challenges in scaling organizational capacity, and internal and external inefficiencies that affect the reliability and effectiveness of cloud services. This paper presents a detailed framework for QA in cloud computing systems and advocates for standardized, automated, and adaptable systems to address these challenges. It aims to establish generic QA policies, incorporate intelligent techniques to enhance extendibility, and create adaptive solutions to manage the inherent attributes of cloud computing environments. The proposed framework is evaluated through survey questionnaires from industry practitioners, and descriptive statistics summarize the results. The study demonstrates the promise, effectiveness, and potential applicability of integrating a single QA framework to enhance the software's functionality, dependability, and future adaptability in cloud computing systems.*

*KEYWORDS*

*Cloud computing, quality assurance, automated systems, adaptive solutions, industry practitioners*

## 1. INTRODUCTION

Cloud computing is a relatively young technology that has drastically changed the technological landscape by offering various organizations advanced and flexible approaches to managing their processes. However, this flexibility also creates numerous challenges in achieving high quality and reliability in software product requirements. Traditional QA methods, designed for static, fixed system architectures, fail to adequately address the decentralized, scale-out nature of cloud solutions [1]. These deficiencies lead to poor services, unmanageability, and negative impacts on system scalability and accuracy, especially where resource scaling is extremely rapid.

There is no universal QA framework, which exacerbates the problems mentioned above. While QA in cloud computing is critical, the practice offers little in terms of performance consistency, stability, and usability because organizations implement various QA approaches [2]. Furthermore, the reduction of centralized control leads to a high dependence on manual actions, resulting in inefficiencies and error sensitivity that diminish the scope of QA in a constantly evolving cloud environment. A crucial aspect involves enhancing the automation and adaptability of QA practices to address the scale and growing complexity of cloud computing.

As a result, this paper will develop a comprehensive framework specifically for QA in the cloud. The framework focuses on three core objectives: integration to ensure consistency in various organizations' QA practices through standardization, scalability through automation, and flexibility to adapt to the ever-evolving nature of cloud systems. These goals aim to be achieved by a framework capable of overcoming the challenges of cloud QA processes and improving software performance. The paper demonstrates the viability and efficiency of the proposed framework through feedback analysis, explaining how its implementation can radically transform the QA processes of the cloud computing industry.

The paper is organized to initially examine the existing literature to identify the potential problems in the existing studies (Section 2). It subsequently outlines the specific challenges associated with cloud-based software (Section 3), and then introduces a novel framework to address the potential challenges that are faced by cloud-based software (Section 4). The validation of the proposed framework is addressed in the section 5.

## 2. RELATED WORK

Ali & Sadeghi [1] found that in today's cloud services landscape, different deployment models-public, private, community, and hybrid—each offer unique benefits and challenges. Migrating sensitive data to cloud infrastructures exposes businesses to greater security threats, such as data deletion, leakages, and vulnerabilities within cloud applications like Google Drive, Dropbox, and Amazon S3. Recent research highlights Data Loss Prevention (DLP) as the most crucial preventive measure against these risks, employing approaches like machine learning, keyword filtering, and Cloud Access Security Brokers (CASB). Specific tools, including CryptDB, Mylar, and Virtru, are developed with unique security solutions in mind, though they also present challenges related to compatibility and platform integration. Additionally, typical security risk evaluation frameworks like OCTAVE Allegro, COBIT 5, and CORAS help model frameworks to assess and mitigate risks within the cloud context, focusing on the confidentiality, integrity, and availability of data. Despite the sophisticated guidance provided by broad frameworks such as ISO 27005 and NIST SP 800-30, there is enough coverage on cloud risks to guide future quality assurance efforts in cloud technologies, which must continue to evolve in response to the ever-growing threats in the cloud computing environment [1].

Environments in cloud infrastructure are dynamic and nearly infinitely scalable. Previous studies have addressed various aspects of quality assurance (QA), including security, performance, and service quality [2]. The flexibility of cloud sharing underscores the need for robust QA mechanisms, as issues like data leakages and service disruptions are likely. The paper highlights the importance of automated testing, continuous integration, and monitoring as essential activities to prevent quality degradation in cloud applications. Moreover, QA for software in the cloud is complicated by the distributed systems, making it challenging to validate requirement testing across different virtualization environments. Quality assurance processes are further complicated by multi-tenancy, where different users share the same resources, making it difficult to identify performance issues or security risks. Various strategies have been proposed to improve QA efficiency in cloud contexts, emphasizing flexibility, scalability, and resource usage. The continuous growth of cloud services and the integration of advanced features like machine learning and artificial intelligence are expected to enhance the accuracy of QA in predicting maintenance and automated faults. However, challenges remain, particularly in aligning quality requirements with customer demand and emerging legislation, as researched by Atoum et al., in 2021 [2].

Ayinla et al. [3] reported on QA frameworks and processes for improving service quality and user experiences in the cloud, emphasizing the importance of organizing Quality Assurance to enhance the implementation and management of cloud computing. They pointed towards the necessity to integrate quality measurement and quality standards in providing cloud services, acting as a guarantee of compliance with required regulations by users of cloud services. The constant assessment of processes and feedback is crucial for maintaining a high standard of services, with research suggesting the use of technological tools to sample and provide real-time feedback on performances. Authors have presented frameworks for measuring the quality of cloud services where user feedback and SLAs are elements of the QA procedures. Hansen also mentioned the application of ISO standards and Frameworks like ITIL as significant in determining the best practices for quality management in the cloud computing environment. As organizations incorporate cloud technologies into their operations, quality assurance plays a distinct role in managing risks typical of data leaks and service disruptions. Integrating these works enhances the understanding of QA in cloud computing and its importance in optimizing operations, improving service delivery, and building trust among cloud service users. Therefore, the present work argues for the applicability and relevance of comprehensive methods of quality assurance for organizations aiming to harness the benefits of cloud computing while being aware of these challenges [3].

QA in cloud computing has become a crucial research area due to the increasing trend of moving operations to the cloud. Significant studies in this field have highlighted the need for high-quality QA methods to ensure the dependability, efficiency, and security of cloud services [4]. Bernardo et al. (2024) explored Software Quality Assurance as a Service, defined by the evaluation of software and cloud services. This model addresses the challenges of quality management in a distributed environment. Key components of QA include continuous supervision, the use of testing tools, and incorporating customer feedback to improve service delivery and reduce operational risks. Additionally, several frameworks provide specific guidelines on organizing QA processes in the context of cloud computing's rapidly evolving environment and culture, including popular methodologies like Agile and DevOps. These frameworks promote a shared responsibility model among development teams, QA teams, and cloud service providers. New technologies such as artificial intelligence and machine learning are also utilized in the quality assessment system to enhance detection capabilities [4].

Calatrava et al. [5] revealed that QA must address features such as authentication, data execution, workload regulation, and resource utilization. Federated Authentication and Authorization Infrastructure (AAI) using institutional Identity Providers (IdPs) remains crucial for security and user convenience. Data management must adhere to FAIR principles, ensuring data is well-named, backed up for long-term access, and searchable through globally recognizable metadata indexers. Effective systems for workload management, such as batch queues, container orchestration, and workflow engines, enable cloud services to manage computationally intensive tasks successfully. Resource management in cloud infrastructures benefits from Infrastructure as Code (IaC), which facilitates the construction and initialization of resources. The quality assurance needs for EOSC Marketplace are met by facilities like EGI Check-in and B2Access-authentication services, and EGI DataHub-data preservation solutions like SLURM and Kubernetes-workload management. The expectations of EOSC regarding information services from the cloud include increases in computer storage space and software quality, and collaboration among various sectors to share practices. These advances provide the stability, security, and scalability necessary for large-scale scientific initiatives [5].

Dawood et al. compiled the approaches to quality assurance in cloud computing with a focus on hybrid settings, revealing that performance and security issues are closely intertwined [6]. As cloud computing grows, the need to ensure quality meets higher standards, where clouds can be classified into models like public, private, community, and hybrid, each with its own unique properties. Key considerations such as data categorization, compatibility issues between on-premise and cloud domains, and successful Identity and Access Management are important for protecting data from loss or exposure during any switch from on premise to cloud. The complexity of compliance policies and resources needed makes it more challenging to implement and manage security events across multiple platforms. A proper approach towards security must include elements like end-to-end encryption and forced monitoring of suspicious activity when multiple clients are allowed to use the same network infrastructure. Moreover, logging and monitoring systems contribute significantly to the identification of threats and the strengthening of cloud infrastructure security. As the study shows, organizations need to design a comprehensive approach to quality assurance that combines security measures with business outcomes to manage cloud computing problems efficiently. This work not only contributes to the understanding of quality assurance of security in cloud computing but also provides directions for further study in evolving mechanisms to cope with unresolved issues. Such a perspective will help organizations improve their cloud initiatives and protect the organization's data and operations' reliability, confidentiality, and accessibility in the rapidly evolving environment [6].

In recent years, various studies have proposed different approaches to assess the quality of cloud services, including contractual service-level agreements that specify performance and reliability [7]. For instance, Dima et al. emphasized that knowledge area analysis is significant in e-learning systems that incorporate cloud computing, highlighting the role of quality assurance frameworks in enhancing the educational process by ensuring effective and efficient service provision. Automated testing and monitoring tools are deemed essential for maintaining quality throughout the service lifecycle, detecting issues before they reach users. Researchers have also stressed the importance of security policy as a critical factor for quality assurance, as robust information security is necessary to attract and retain cloud service clients. Given the rapid development of cloud technologies, quality assurance standards must stay relevant to current developments and be updated continually. Proper implementation of quality assurance can increase user satisfaction and support the development of cloud computing services across various sectors such as education, business, and healthcare, as organizations increasingly focus on cloud computing, underscoring the critical need for effective and secure quality assurance [7].

Kedi et al. [8] reviewed data storage and analysis solutions for healthcare, emphasizing data security, compliance, and availability. As healthcare systems increasingly migrate to the cloud, concerns such as data vulnerability to cyber-attacks and compliance with legal requirements become prominent. The integration of artificial intelligence (AI), machine learning (ML), and the Internet of Things (IoT) into cloud services has the potential to add value through data analysis, improved patient care, and optimized business processes. Blockchain technology also holds promise for maintaining data integrity and security, which is crucial for QA in cloud computing. However, the need to design effective disaster recovery solutions and integrate them with existing systems highlights the importance of adequate planning and risk management in healthcare. The constant advancement in cloud computing presents opportunities for enhancing healthcare delivery, but these must be accompanied by deliberate efforts to mitigate quality assurance risks. Ensuring quality in cloud services not only protects patients' data but also ensures better healthcare solutions through technological enhancement and efficient techniques, underlining the importance of quality control in cloud organization [8].

The assurance of quality in cloud computing has evolved, particularly in manufacturing, where operations are increasingly automated to maintain the standard of products [9-10]. Previously, visual and implicit inspections were common, but now, Artificial Intelligence (AI) and Machine Learning (ML) are employed to enhance task performance and productivity. Heteronomous systems, such as equipment calibration, predictive maintenance, and real-time quality inspections, are now run through cloud-based computing systems. These systems allow manufacturing facilities to shift computationally intensive operations to cloud servers or edge clouds, reducing latency and enhancing decision-making speed. Visual inspection systems utilize the cloud to process real-time images of products to identify defects. This methodology not only improves the speed and efficiency of inspections but also reduces the necessity for centralized computing structures at each station, making it a more scalable process. Moreover, the success of implementing computer vision along with AI has enabled models like ResNet50, used in some studies, to enhance the defect detection rate up to 93%, surpassing previous methods. Future efforts will focus on minimizing response time and improving the communication interface between cloud servers and onsite sensors. The advancement and increased application of cloud computing with AI in quality control are found to provide scalable and enhanced operational effectiveness for today's industrial complexes, as identified [9-10].

Table 1 displays a concise overview of the limitations identified in the related work.

Table 1. The main limitations of related work.

| Title | Limitations |
| --- | --- |
| Information Security Risk Assessment Methods in Cloud Computing: Comprehensive Review [1] | Reactive risk assessment methods fail to predict emerging threats, limiting proactive security measures. |
| Challenges of Software Requirements Quality Assurance and Validation: A Systematic Literature Review [2] | Lack of standardized methodologies for validating cloud-based systems, affecting scalability and consistency. |
| Enhancing accounting operations through cloud computing: A review and implementation guide [3] | Limited infrastructure and regulatory frameworks delay the adoption of cloud-based solutions in accounting, particularly in certain regions. |
| Software Quality Assurance as a Service: Encompassing the quality assessment of software and services [4] | Fragmented QA processes without holistic integration, limiting automation effectiveness. |
| A survey of the European Open Science Cloud services for expanding the capacity and capabilities of multidisciplinary scientific applications [5] | Interoperability issues between different cloud platforms hinder cross-disciplinary research and limit scalability. |
| Cyberattacks and Security of Cloud Computing: A Complete Guideline [6] | Slow response times in detecting and mitigating sophisticated cyberattacks, threatening system integrity. |
| Mapping Knowledge Area Analysis in E-Learning Systems Based on Cloud Computing [7] | Cloud solutions often fail to provide personalized user experiences, reducing effectiveness in diverse educational settings. |
| Cloud computing in healthcare: A comprehensive review of data storage and analysis solutions [8] | Real-time data availability challenged by latency and bandwidth issues, particularly in remote areas. |
| IT standardization in cloud computing: Security challenges, benefits, and future directions [9] | Inadequate international collaboration on security standards, increasing the risk of data breaches. |
| An Improved Intelligent Cloud-based Structure for Automated Product Quality Control [10] | Communication delays between on-site equipment and cloud servers reduce the efficiency of real-time monitoring in production environments. |

## 3. PROBLEM DEFINITION

Atoum et al. [2] identify a research gap in cloud computing concerning the absence of a comprehensive model for maintaining the quality of software requirements QA and validation in cloud environments. These methods exhibit usability deficiencies as they vary from one company to another, resulting in performance, stability, and scalability differences in the software. This issue is exacerbated in cloud environments due to their dynamic and distributed nature, which conventional measures often fail to adequately address. Furthermore, no recording formats are standardized across the business environment, leading to unstructured QA processes that necessitate many manual actions and are highly prone to errors. These disparities complicate the scaling and automation of cloud services, reducing their reliability, especially in many emerging cloud systems. A primary concern highlighted in the paper is the inflexibility of extending existing QA frameworks for use in cloud computing environments, attributed to the technology's elastic and on-demand nature. Some organizations face challenges in maintaining quality services when cloud resources are automatically scaled up or down, leading to unpredictable service quality. Consequently, end-users experience a significant decline in service performance, particularly when the load increases.

Following is a research question based on the problem discussed in [2].

- How can a standardized quality assurance framework be developed to address the dynamic and distributed nature of cloud computing environments and ensure consistency in software performance and scalability?

This question seeks to identify how an enhanced QA framework for cloud computing applications can be developed to accommodate more flexible and dynamic requirements, thereby reducing disparities and frequent interventions.

## 4. THE PROPOSED SOLUTION

To address the research question of developing a standardized quality assurance (QA) framework that can adapt to the dynamic and distributed nature of cloud computing environments while ensuring consistency in software performance and scalability, a novel QA framework is proposed. This framework integrates three goals i.e., standardization, automation, and adaptability, tailored specifically to meet the unique challenges of cloud computing as shown in figure 1.

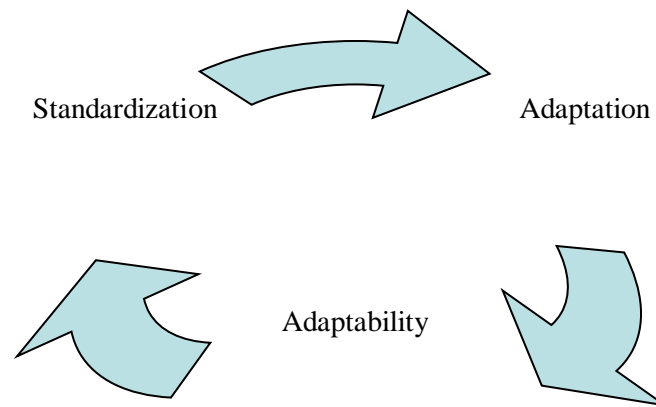

Figure 1. The proposed framework

### 4.1. Goal 1: Standardization of QA Practices Across Cloud Platforms

The first pillar of the proposed framework is the standardization of QA practices. This involves creating a set of universally applicable QA guidelines and procedures that are adopted across different organizations and platforms. The aim is to mitigate variability in QA practices that often leads to inconsistencies in software performance and scalability. To achieve this, the framework would:

- Develop standardized testing protocols and performance benchmarks tailored for cloud computing.
- Facilitate a common understanding and implementation of QA standards through collaboration with industry leaders, cloud service providers, and regulatory bodies.
- Introduce certification processes for cloud applications and platforms that comply with these standards, ensuring a baseline quality level across the industry.

### 4.2. Goal 2. Automation of QA Processes

Automation is crucial in addressing the scalability and dynamic resource allocation inherent to cloud environments. By automating QA processes, the framework can provide real-time monitoring and testing, which is essential for maintaining performance standards despite the fluid nature of cloud resource utilization. The automation strategy would include:

- Deployment of automated testing tools that can dynamically scale with the cloud resources, capable of executing stress tests and performance evaluations in real time.
- Integration of continuous integration and continuous deployment (CI/CD) pipelines with automated QA checks to ensure that every update or change adheres to the defined QA standards.
- Utilization of machine learning algorithms to predict potential breakdowns and bottlenecks based on historical data, enhancing the proactive capabilities of the QA framework.

### 4.3. Goal 3. Adaptability to Cloud-Specific

Challenges The third component focuses on the adaptability of the QA framework. Given the diverse and often unpredictable nature of cloud environments, the framework must be capable of adjusting its operations to different scenarios and conditions. This adaptability can be achieved through:
- Implementing flexible QA frameworks that can be customized for different types of cloud services and deployment models (e.g., public, private, hybrid).
- Developing adaptive testing methods that adjust their complexity and scope based on the current load and performance data.
- Establishing feedback loops within the QA process that allow continuous learning and improvement based on real-time performance data and user feedback.

To implement this novel QA framework, a phased approach can be adopted as follows.

Phase 1: Conduct a pilot project with select cloud providers to refine the standardization protocols and automate QA processes.
Phase 2: Roll out the standardized QA procedures across a wider range of cloud services and platforms, integrating feedback from the pilot to improve the processes.
Phase 3: Fully integrate adaptability features into the QA framework, ensuring that it can respond effectively to the dynamic nature of cloud computing environments.

The anticipated outcome of this novel QA framework includes:

- Higher consistency in software performance across different cloud platforms.
- Improved scalability of cloud services with minimal human intervention.
- Enhanced reliability and user trust in cloud-based applications and services.

By addressing these three goals (standardization, automation, and adaptability), the proposed QA framework aims to significantly enhance the quality assurance of cloud computing environments, making them more robust, scalable, and consistent. A sequence diagram is used to show the functional representation of the proposed framework in Figure 2.

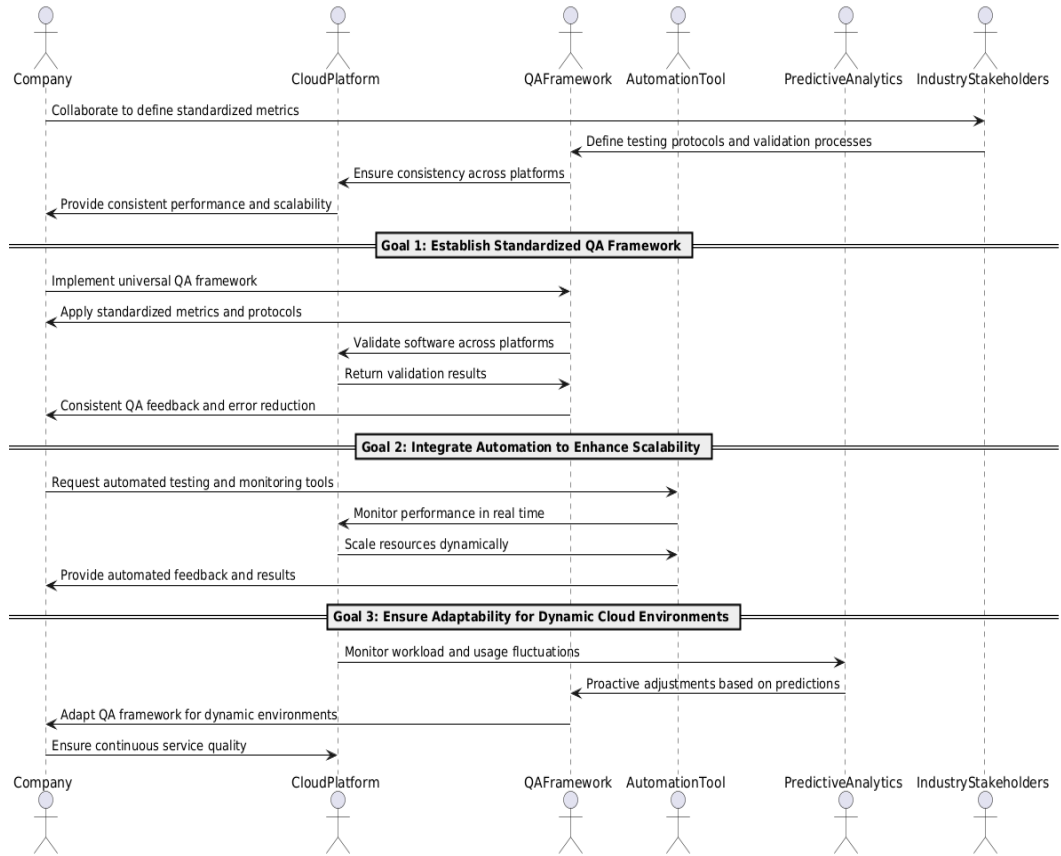

Figure 2. The functional representation of the proposed framework using a squence diagram

## 5. VALIDATION OF THE PROPOSED SOLUTION

The sample size is critical in quantitative research, particularly when the researcher aims to understand respondents' impressions, attitudes, and feedback regarding a specific solution. Questionnaires, structured with limited predetermined answer choices, are commonly used. These surveys are effective tools for assessing reactions and various factors influencing the research subject. This study utilizes a structured questionnaire to evaluate the Quality Assurance (QA) framework for cloud computing. Thirty respondents from various cloud computing companies were selected using a random sampling technique suitable for this type of research. The target population of thirty is adequate for calculating the mean, variance, and cumulative frequency from the observed responses. This research aims to assess the feasibility, versatility, and seamless integration of the proposed QA framework into cloud environments. The questionnaire explores how well the proposed framework addresses issues such as service quality, automation, and dynamic system changes. Data is presented using frequency tables and bar charts, while descriptive analysis is used to analyse the data.

n/aInternational Journal of Software Engineering & Applications (IJSEA), Vol.16, No.1, January 2025

## 5.1. Goal 1 - Establish a Standardized QA Framework

The questions under Goal 1 are aimed at evaluating the effectiveness of the proposed standardized QA framework to ensure that quality standards are maintained across different processes. Table 2 demonstrates a diverse range of responses across five different levels of satisfaction, from 'Very Low' to 'Very High'. Table 2 shows that a large portion of the respondents, 37.30% on average, find the current QA framework nominal, indicating a moderate level of acceptance. This is followed closely by 35.14% of the respondents who rated the framework as 'High', suggesting a significant inclination towards a more rigorous QA process. Furthermore, 16.76% felt very highly about the framework, reflecting a strong endorsement of robust quality measures. However, not all feedback was positive. On average, 7.03% of the participants rated the framework as 'Low', showing some level of dissatisfaction. A smaller percentage, 3.78%, viewed the QA framework as 'Very Low', expressing strong discontent with the current standards.

Table 2. Cumulative analysis of Goal 1

| Goal 1 | | | | | |
|---|---|---|---|---|---|
| Q. No. | Very Low | Low | Nominal | High | Very High |
| Q1 | 2.70 | 5.41 | 45.95 | 27.03 | 18.92 |
| Q2 | 0.00 | 2.70 | 37.84 | 35.14 | 24.32 |
| Q3 | 10.81 | 5.41 | 32.43 | 32.43 | 18.92 |
| Q4 | 0.00 | 10.81 | 35.14 | 45.95 | 8.11 |
| Q5 | 5.41 | 10.81 | 35.14 | 35.14 | 13.51 |
| Total | 18.92 | 35.14 | 186.49 | 175.68 | 83.78 |
| Avg. | 3.78% | 7.03% | 37.30% | 35.14% | 16.76% |

## 5.2. Goal 2 - Integrate Automation to Enhance Scalability

In the cumulative analysis presented in Table 3, a minimal 8.11% of total responses, with an average of 1.62%, fall into the 'Very Low' category, indicating that extreme dissatisfaction with the system integrations is rare and does not highlight significant fundamental issues. A slightly higher response, totalling 32.43% and averaging 6.49%, is observed in the 'Low' satisfaction level. This feedback, while not predominant, pinpoints specific areas where the integrations could be significantly improved to meet stakeholder expectations. The majority of stakeholders, reflected in a 200.00% total response rate and a 40.00% average, categorize their satisfaction as 'Nominal'. This indicates a general acceptance of the system integrations as adequate, suggesting that while the essentials are met, there is potential to enhance these integrations to achieve higher satisfaction. Responses in the 'High' satisfaction level, which total 162.16% and average 32.43%, demonstrate that many aspects of the system integrations are effective and well-received, indicating that these elements are meeting or exceeding the needs of stakeholders. This positive feedback suggests that these aspects can serve as models for refining other parts of the system.

Furthermore, nearly one-fifth of the feedback, with a total of 97.30% and an average of 19.46%, is extremely positive, ranking the system integrations as 'Very High'. This level of satisfaction underscores that certain elements of the integration are exemplary, highlighting them as benchmarks for ongoing improvements across the integration spectrum. This analysis illustrates that the system integrations generally meet stakeholder needs effectively, but there is a distinct opportunity to elevate satisfaction from nominal to higher levels. Such enhancements could maximize the effectiveness of the QA framework, with the feedback largely constructive and skewed towards more positive responses, allowing for targeted improvements in areas of lower satisfaction.

Table 3. Cumulative analysis of Goal 2

| Goal 2 | | | | | |
|---|---|---|---|---|---|
| Q. No. | Very Low | Low | Nominal | High | Very High |
| Q1 | 5.41 | 5.41 | 40.54 | 32.43 | 16.22 |
| Q2 | 2.70 | 8.11 | 43.24 | 32.43 | 13.51 |
| Q3 | 0.00 | 5.41 | 43.24 | 37.84 | 13.51 |
| Q4 | 0.00 | 8.11 | 37.84 | 29.73 | 24.32 |
| Q5 | 0.00 | 5.41 | 35.14 | 29.73 | 29.73 |
| Total | 8.11 | 32.43 | 200.00 | 162.16 | 97.30 |
| Avg. | 1.62% | 6.49% | 40.00% | 32.43% | 19.46% |

## 5.3. Goal 3 - Facilitation of Communication and Negotiation

The cumulative analysis of Goal 3, outlined in Table 4, spans a wide range of stakeholder satisfaction levels, from 'Very Low' to 'Very High'. The majority of feedback falls within the 'Nominal' category, with an average of 35.68%, suggesting that while stakeholders generally find the initiatives acceptable, there is significant potential for improvement. At the lower end of the spectrum, the 'Very Low' and 'Low' satisfaction levels register averages of 5.41% and 6.49% respectively. These figures highlight specific areas within Goal 3 that are viewed critically by stakeholders, indicating where improvements are most needed. On the positive side, 'High' satisfaction is reported at an average of 31.89%, and 'Very High' satisfaction at 20.54%, signalling that a substantial portion of the feedback is very favourable. These higher satisfaction levels underscore the success of certain elements of the initiatives, which are well-received and exceed stakeholder expectations. This analysis suggests a generally positive reception but also underscores the importance of enhancing areas that currently achieve only nominal satisfaction. By focusing on these areas, the initiatives under Goal 3 could potentially elevate overall satisfaction, leveraging the strong aspects to improve the weaker ones and better align with stakeholder expectations.

Table 4. Cumulative analysis of Goal 3

| Goal 3 | | | | | |
|---|---|---|---|---|---|
| Q. No. | Very Low | Low | Nominal | High | Very High |
| Q1 | 2.70 | 13.51 | 45.95 | 18.92 | 18.92 |
| Q2 | 0.00 | 5.41 | 40.54 | 35.14 | 18.92 |
| Q3 | 8.11 | 5.41 | 37.84 | 24.32 | 24.32 |
| Q4 | 2.70 | 2.70 | 24.32 | 45.95 | 24.32 |
| Q5 | 13.51 | 5.41 | 29.73 | 35.14 | 16.22 |
| Total | 27.03 | 32.43 | 178.38 | 159.46 | 102.70 |
| Avg. | 5.41% | 6.49% | 35.68% | 31.89% | 20.54% |

International Journal of Software Engineering & Applications (IJSEA), Vol.16, No.1, January 2025

Table 5. Final Cumulative Analysis of Three Gaols

| Goal No. | Very Low | Low | Nominal | High | Very High |
|---|---|---|---|---|---|
| 1 | 3.78% | 7.03% | 37.30% | 35.14% | 16.76% |
| 2 | 1.62% | 6.49% | 40.00% | 32.43% | 19.46% |
| 3 | 5.41% | 6.49% | 35.68% | 31.89% | 20.54% |
| Total | 10.81% | 20.00% | 112.97% | 99.46% | 56.76% |
| Avg. | 3.60% | 6.67% | 37.66% | 33.15% | 18.92% |

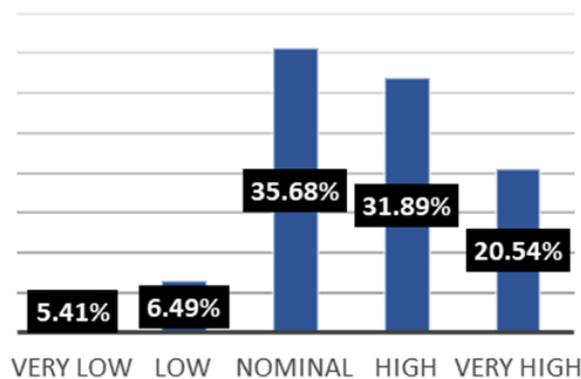

Figure 3. Final cumulative analysis of three goals

## 5.4. Final Cumulative Analysis of Three Goals

The cumulative analysis from Table 5 shows stakeholder satisfaction across three goals, highlighting general acceptance with an opportunity for enhancement. The very low and low satisfaction levels are minimal, averaging 3.60% and 6.67% respectively, suggesting only minor fundamental issues across the initiatives. Most feedback falls into the 'Nominal' category with an average of 37.66%, indicating that while stakeholders find the initiatives adequate, there's significant room for improvement. The 'High' and 'Very High' satisfaction levels, averaging 33.15% and 18.92% respectively, show that many aspects of the initiatives are well-received, particularly in Goals 2 and 3, which are seen as exemplary. This positive feedback suggests effective elements that could be leveraged to enhance lesser-performing areas. Overall, the data indicates that elevating nominal satisfaction could substantially increase the effectiveness and overall approval of these initiatives as shown in figure 3.

## 6. CONCLUSIONS AND FUTURE WORK

The self-assessment reveals a promising start and substantial progress toward achieving three critical objectives. While broad successes have been noted, it is crucial to implement specific practices aimed at addressing particular issues and enhancing underperforming areas. As these objectives evolve and expand, optimizing technology use and fostering greater collaboration across departments will be key to the sustainable success of the planned initiatives. To build on current achievements and address existing gaps, several strategies are proposed. First, enhance standardization efforts by improving training and documentation, which supports the uniform application of the QA framework across all teams. Second, refine automation strategies by identifying barriers to scalability and incorporating machine learning techniques to increase capacity. Third, improve cloud adaptability by adopting predictive analytics and other AI-driven solutions to manage the dynamic needs of cloud environments effectively. Fourth, address areas with low scores by developing targeted resources and improvement plans to ensure uniform performance across all metrics. Lastly, promote cross-goal synergy by encouraging integration between the frameworks for standardization, scaling capacity, and adaptability through the design of sophisticated tools.

To further develop the proposed quality assurance (QA) framework for cloud-based software, future work could focus on several key areas. Advanced automation techniques, especially those leveraging AI and ML, could be used to enhance predictive capabilities and proactive issue identification. Expanding validation studies across different cloud environments and industries would help test the framework's robustness and general applicability. Integrating the QA framework with CI/CD pipelines would improve its adaptability to rapid development cycles and changes in cloud environments. Addressing regulatory compliance and security within the framework could cater to growing concerns in these areas, and customizing the framework for specific industries would allow it to meet unique challenges effectively. Incorporating real-time user feedback would enable ongoing adjustments based on actual use, while ensuring cross-platform compatibility would make the framework versatile across different cloud technologies. Finally, developing comprehensive educational and training programs would support organizations in effectively adopting and operationalizing the QA framework, enhancing its practical utility and impact.

International Journal of Software Engineering & Applications (IJSEA), Vol.16, No.1, January 2025

## AUTHORS

**Prof. Dr. Rizwan Qureshi** received his Ph.D. degree in Computer Sciences from National College of Business Administration & Economics, Pakistan 2009. He is currently working as a Professor in the Department of IT, King Abdulaziz University, Jeddah, Saudi Arabia. This author is the best researcher awardees from the Department of Information Technology, King Abdulaziz University in 2013 and 2016.

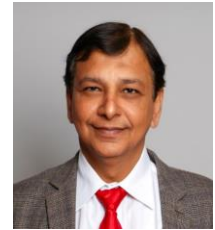

**Mr. Mohammed Ahmad Alharbi** is a master student in the Department of IT, King Abdulaziz University, Jeddah, Saudi Arabia.